\documentclass[aps,prl,twocolumn,showpacs]{revtex4}
\usepackage{dcolumn}
\usepackage{amsmath,amsthm,amssymb}
\usepackage{latexsym}
\usepackage{array}
\usepackage{bm,textcomp}
\begin{document}
Europhys. Lett. {\bf 83}, 50002 (2008)
\title{Geometric vs. Dynamical Gates in
Quantum Computing Implementations Using Zeeman and Heisenberg
Hamiltonians}
\author{Yu Shi}
\thanks{Email: yushi@fudan.edu.cn}
\affiliation{Department of Physics, Fudan University, Shanghai
200433, China}
\begin{abstract}
Quantum computing in terms of geometric phases, i.e. Berry or
Aharonov-Anandan phases, is fault-tolerant to a certain degree. We
examine its implementation based on Zeeman coupling with a rotating
field and isotropic Heisenberg interaction, which describe NMR and
can also be realized in quantum dots and cold atoms. Using a novel
physical representation of the qubit basis states, we construct
$\pi/8$ and Hadamard  gates based on Berry and Aharonov-Anandan
phases. For two interacting qubits in a rotating field, we find that
it is always impossible to construct a two-qubit gate based on Berry
phases, or based on Aharonov-Anandan phases when the gyromagnetic
ratios of the two qubits are equal. In implementing a universal set
of quantum gates, one may combine geometric $\pi/8$ and Hadamard
gates and dynamical $\sqrt{\rm SWAP}$ gate.

\end{abstract}
 \maketitle

Berry phase is the geometric phase accumulated in a cycle of the
adiabatically varying parameters of the Hamiltonian, depending only
on the path in the parameter space~\cite{berry}.  Aharonov-Anandan
phase, for which adiabaticity is not necessary, is the geometric
phase accumulated in a cycle of the state itself, depending only on
the path in the projected Hilbert space of rays~\cite{aharonov}.
Insensitive to dynamical details, they have been exploited as a
hardware approach to fault-tolerant quantum
computing~\cite{jones,zhu,wang,leibfried}. The basic idea is the
following. For one or two qubits, one finds a set of basis states
$|\psi_i\rangle$, each of which is designed to transform only by a
geometric phase $\gamma^g_i$: $|\psi_i\rangle \rightarrow
e^{i\gamma_i^g}|\psi_i\rangle$. Then an arbitrary state transforms
as $\sum a_i|\psi_i\rangle \rightarrow \sum a_i
e^{i\gamma^g_i}|\psi_i\rangle$. Consequently quantum gates based on
geometric phases may be constructed.  It has been demonstrated
theoretically~\cite{chiara} and confirmed experimentally~\cite{leek}
that geometric phases are indeed resistant to some types of noise,
error and decoherence.  A quantum gate based on Berry phase must
satisfy the adiabatic condition that the gating time is much longer
than $\hbar$ divided by the minimal energy difference between the
qubit basis state and other energy eigenstates. This is not needed
for a quantum gate based on Aharonov-Annandan phase, hence it is
more advantageous from the perspectives of both shortening computing
time and overcoming decoherence~\cite{zhu}.

Let us focus on the implementations using Zeeman coupling with a
rotating field and isotropic Heisenberg interaction. Zeeman coupling
effectively describes probably all implementations of single-qubit
gates. Isotropic Heisenberg interaction describes effective coupling
between nuclear spins mediated by electrons on a chemical bond in
liquid NMR~\cite{slichter}.  Recent experimental advances in
coherent control in quantum dots~\cite{petta} and in cold
atoms~\cite{atoms,bloch} make it interesting to consider geometric
quantum computing in these scalable systems, in which the
interaction is Heisenberg-type as well. In cold atoms in an optical
lattice,  the Heisenberg interaction is due to either direct
exchange energy~\cite{atoms}, or interplay between tunneling and
interaction~\cite{bloch}, similar to the well known case of double
quantum dots~\cite{loss}. Excellent pioneering discussions and an
NMR experimental demonstration have been made on two-qubit geometric
phases with one of the two qubits regarded as decoupled with the
rotating field~\cite{jones,zhu,wang}. These analyses were based on
Ising-type interaction, which can be realized in Josephson junctions
and in cold atoms, while in liquid NMR, it is a good approximation
when the Heisenberg interaction is weak or when the two spins have
vastly different precession frequencies~\cite{nielsen}. This
reasonable approximation nicely simplifies the calculations and
experimental designs, and has been successful. Nevertheless, it
remains to be examined whether nontrivial two-qubit geometric phases
can arise at all under the isotropic Heisenberg interaction and with
both qubits coupled with the field. In this Letter, we study the
Berry and Aharonov-Anandan phases of two Heisenberg-coupled qubits,
both of which are coupled with the rotating field. For Berry phases,
and for Aharonov-Anandan phases in case the gyromagnetic ratios of
the two qubits are equal, the geometric part of the unitary
transformation is always trivial, i.e. it is  a product of two
single-qubit transformation, rendering it impossible to construct
any nontrivial two-qubit geometric gate. For Aharonov-Anandan phases
in case the gyromagnetic ratios of the two qubits are different, the
geometric part of the two-qubit transformation can be nontrivial, in
consistency with the previous results. Before discussing two-qubit
operations, we construct $\pi/8$ and Hadamard gates based on Berry
or Aharonov-Anandan phases by using a novel representation of the
qubit basis states. Previously, in the geometric implementations of
quantum computing using these Hamiltonians, there has been a lack of
explicit construction of a standard universal set of quantum gates
such as the one consisting of $\pi/8$, Hadamard and controlled-NOT
gates. This is an important issue, because although there were
existence proofs that most two-qubit gates are universal while two
non-commuting single-qubit gates can generate an arbitrary one-qubit
transformation~\cite{deutsch}, it is generically unknown how to
actually realize an arbitrary operation.

Let us start with a single spin-$\frac{1}{2}$ in a magnetic field,
with  the Hamiltonian
\begin{equation}
h=-{\kappa}[s_zB_0+s_xB_1\cos\phi(t)+s_yB_1\sin\phi(t)]=
 -\kappa\hat{\mathbf{s}}\cdot\mathbf{B}(t),
\label{single}
\end{equation}
where $\kappa$ is  the gyromagnetic ratio, $\hbar=1$,
$\mathbf{B}(t)=[B_1\cos \phi(t), B_1\sin \phi(t), B_0]$ is the total
field, where $B_0$ is the static component along $z$ axis, $B_1$ is
the magnitude of the rotating component on $xy$ plane, $\phi(t)$ is
the rotating angle, $\hat{\mathbf{s}}=(\frac{\hat{\sigma}_x}{2},
\frac{\hat{\sigma}_y}{2}, \frac{\hat{\sigma}_z}{2})$, $\sigma$'s
being  Pauli operators. The Hamiltonian can be rewritten as
$h=-\kappa\hat{\sigma}_{\mathbf{B}(t)} B/2$, where
$\hat{\sigma}_{\mathbf{B}(t)}$ is the Pauli operator along the
direction of $\mathbf{B}(t)$, $B=\sqrt{B_0^2+B_1^2}$ is the
magnitude of the total field and is time-independent.

First consider the adiabatic limit, in which  the functional form of
$\phi(t)$ can even be arbitrary as far as it reaches $2\pi$ after a
period $\tau$. The instantaneous eigenstates of $h$ are just the
instantaneous eigenstates $|\sigma_{\mathbf{B}(t)}\rangle$ of
$\hat{\sigma}_{\mathbf{B}(t)}$: $
|1_{\mathbf{B}(t)}\rangle \equiv |\uparrow_{\mathbf{B}(t)}\rangle =
 \cos\frac{\theta}{2}|\uparrow_z\rangle+
\sin\frac{\theta}{2}e^{i\phi(t)}|\downarrow_z\rangle,$
$|-1_{\mathbf{B}(t)}\rangle \equiv
|\downarrow_{\mathbf{B}(t)}\rangle  =
\sin\frac{\theta}{2}|\uparrow_z\rangle-
\cos\frac{\theta}{2}e^{i\phi(t)}|\downarrow_z\rangle,
$
where $\theta=\arctan \frac{B_1}{B_0}$.

Suppose at $t=0$,  the  state is $|\sigma_{\mathbf{B}(0)}\rangle$,
where $\sigma_{\mathbf{B}(0)}=\pm 1$, representing $\uparrow$ or
$\downarrow$ along the direction of $\mathbf{B}(0)$. After a cycle,
the state becomes $e^{i(\gamma^d_{\sigma_{\mathbf{B}(0)}}
+\gamma^b_{\sigma_{\mathbf{B}(0)}})}|\sigma_{\mathbf{B}(0)}\rangle$,
where
$\gamma^d_{\sigma_{\mathbf{B}(0)}}=\kappa\sigma_{\mathbf{B}(0)} B
\tau/2$ is the dynamical phase, while
$\gamma^b_{\sigma_{\mathbf{B}(0)}}=-\sigma_{\mathbf{B}(0)}\Omega(\theta)/2$
is the Berry phase, where $\Omega(\theta)=2\pi(1-\cos\theta)$  is
the solid angle that the circuit of $\mathbf{B}(t)$ subtends at
$\mathbf{B}=0$. If the initial state is an arbitrary superposition
$\alpha_{\uparrow}|\uparrow_{\mathbf{B}(0)}\rangle +
\alpha_{\downarrow}|\downarrow_{\mathbf{B}(0)}\rangle$, then after a
cycle, the state becomes $\alpha_{\uparrow}
e^{i(\gamma^d_{\uparrow}+\gamma^b_{\uparrow})}|\uparrow_{\mathbf{B}(0)}\rangle
+ \alpha_{\downarrow}e^{i(\gamma^d_{\downarrow}+
\gamma^b_{\downarrow})}|\downarrow_{\mathbf{B}(0)}\rangle$. In other
words, the unitary transformation in a cycle is
$diag(e^{i(\gamma^d_{\downarrow}+\gamma^b_{\downarrow})},
e^{i(\gamma^d_{\downarrow}+\gamma^b_{\downarrow})}),$  written in
the basis
$\{|\uparrow_{\mathbf{B}(0)}\rangle,|\downarrow_{\mathbf{B}(0)}\rangle\}$.

The dynamical phase in each eigenstate can be canceled in two
consecutive cycles with opposite directions of $\mathbf{B}(0)$, i.e.
with both $z$ and $xy$ components of the field reversed at the
beginning of the second cycle. Equivalently, one may reverse the
spin using the method of refocus or spin echoes~\cite{jones}. After
these two cycles,  the Berry phase doubles to
$-\sigma_{\mathbf{B}(0)}\Omega(\theta)=\sigma_{\mathbf{B}}2\pi
B_0/\sqrt{B_0^2+B_1^2}$, up to $2\pi n$ ($n$ is an integer).
Consequently, for an arbitrary initial state, the  unitary
transformation is purely based on Berry phases, and can be written
as
\begin{equation} U_{1,berry}=diag(e^{i\gamma},
e^{-i\gamma}), \label{1b} \end{equation} where $\gamma=2\pi
B_0/\sqrt{B_0^2+B_1^2}$.

We represent the qubit basis state as the spin eigenstates along the
direction $(B_1,0,B_0)$, rather than along $z$ direction, as usually
do. Then it is straightforward that $U_{1,berry}$ becomes $\pi/8$
gate $diag(e^{-i\pi/8}, e^{i\pi/8})$ when $B_0/|B_1|=-1/\sqrt{255}$.

The Hadamard gate $ \frac{1}{\sqrt{2}}\left ( \begin{array}{cc}1 & 1
\\ 1& -1\end{array} \right ) $ is constructed also by using
the Hamiltonian (\ref{single}), but the direction of
$\mathbf{B}(0)$ is rotated by an angle $\chi$ with respect to $y$
axis, from the direction $(B_1,0,B_0)$, which always defines the
qubit basis states. For two cycles with opposite signs of
$\mathbf{B}(0)$, or equivalently, opposite signs of spin as realized
by using spin echoes, one obtains the purely geometric unitary
transformation as given in (\ref{1b}), in the basis
$\{|\uparrow_{\mathbf{B}(0)}\rangle,
|\downarrow_{\mathbf{B}(0)}\rangle\}$ for the new direction of
$\mathbf{B}(0)$. In the qubit basis, it should be written as
\begin{equation} \left (\begin{array}{cc}
e^{i\gamma}\cos^2\frac{\chi}{2}+e^{-i\gamma}\sin^2\frac{\chi}{2}
& i\sin\gamma\sin\chi \\
i\sin\gamma\sin\chi&
e^{i\gamma}\cos^2\frac{\chi}{2}+e^{-i\gamma}\sin^2\frac{\chi}{2}
\end{array}
\right), \label{had} \end{equation}  which  becomes the Hadamard
gate, up to a global factor $i$, when $\chi=\pi/4$, $\gamma=\pi/2$.
This value of $\gamma$ is obtained when $B_0/|B_1|=1/\sqrt{15}$.

Besides the usual robustness of geometric phases~\cite{chiara,leek},
an additional aspect of robustness of the quantum gates constructed
above is that for the two consecutive cycles whose dynamical phases
cancel each other, only the period $\tau$ needs to be the same,
while the time-dependence of $\phi(t)$ can be completely arbitrary
and independent in each cycle. The underlying reason is that the
energy for each instantaneous eigenstate is time-independent, thus
the dynamical phase is simply a product of the energy and the
period.

Now we show how to use Aharonov-Anandan phases to  construct the
$\pi/8$ and Hadamard gates, under the same single-qubit Hamiltonian
(\ref{single}).  No requirement of adiabaticity is needed, but it is
specified that $\phi(t)=\omega t$.  The solution of (\ref{single})
can always be given as~\cite{slichter}
\begin{equation} |\psi(t)\rangle = e^{-is_z \omega t}
e^{-i{\tilde{h}}t}|\psi(0)\rangle, \label{solution} \end{equation}
where
$$\tilde{h}= h(0)-\omega \hat{s}_z =
 -\kappa \hat{\mathbf{s}}\cdot \tilde{\mathbf{B}}$$ is
time-independent, $\tilde{\mathbf{B}}= (B_1, 0, B_0+\omega/\kappa),
$ whose magnitude is $\tilde{B} =
\sqrt{(B_0+\omega/\kappa)^2+B_1^2}$. The eigenstates of $\tilde{h}$
are just the eigenstates $|\sigma_{\tilde{\mathbf{B}}}\rangle$ of
the spin operator $\hat{\sigma}_{\tilde{\mathbf{B}}}$ along the
direction of $\tilde{\mathbf{B}}$, with eigenvalue
$-\kappa\sigma_{\tilde{\mathbf{B}}}\tilde{B}/2$, i.e. $
|1_{\tilde{\mathbf{B}}}\rangle \equiv
|\uparrow_{\tilde{\mathbf{B}}}\rangle  =
\cos\frac{\tilde{\theta}}{2}|\uparrow_z\rangle+
\sin\frac{\tilde{\theta}}{2}|\downarrow_z\rangle,$
$|-1_{\tilde{\mathbf{B}}}\rangle \equiv
 |\downarrow_{\tilde{\mathbf{B}}}\rangle
= \sin\frac{\tilde{\theta}}{2}|\uparrow_z\rangle -
\cos\frac{\tilde{\theta}}{2}|\downarrow_z\rangle,$
where $\tilde{\theta}=\arctan \frac{B_1}{B_0+\frac{\omega}{\kappa}
}.$

If we start with $|\sigma_{\tilde{\mathbf{B}}}\rangle$ (in the
original frame, instead of a ``rotating frame''), then after the
period $\tau=2\pi/\omega$, the state is
$|\psi_{\tilde{\mathbf{B}}}(\tau)\rangle
=e^{i\gamma}|\sigma_{\tilde{\mathbf{B}}}\rangle$, where $\gamma=\pi
+\kappa\sigma_{\tilde{\mathbf{B}}}\tilde{B}\tau/2$ is the total
phase, in which $\kappa\sigma_{\tilde{\mathbf{B}}}\tilde{B}\tau/2$
appears because $|\sigma_{\tilde{\mathbf{B}}}\rangle$ is an
eigenstate of $\tilde{h}$, while $\pi$ appears because
$\hat{s}_z|\sigma_z\rangle = \pm 1/2$, thus any state is an
eigenstate of $e^{-i s_z\omega \tau}$ with eigenvalue $e^{i\pi}$.
The  dynamical phase is $\gamma^d_{\sigma_{\tilde{\mathbf{B}}}} =
-\int dt \langle
\psi_{\tilde{\mathbf{B}}}(t)|h(t)|\psi_{\tilde{\mathbf{B}}}(t)\rangle
= -\tau \langle
\sigma_{\tilde{\mathbf{B}}}|h(0)|\sigma_{\tilde{\mathbf{B}}}\rangle
=-\tau\langle
\sigma_{\tilde{\mathbf{B}}}|\tilde{h}|\sigma_{\tilde{\mathbf{B}}}\rangle+\omega\tau
\langle
\sigma_{\tilde{\mathbf{B}}}|s_z|\sigma_{\tilde{\mathbf{B}}}\rangle
=\sigma_{\tilde{\mathbf{B}}}(\kappa\tilde{B}\tau/2
-\pi\cos\tilde{\theta})$,  while the Aharonov-Anandan phase is
$\gamma^{aa}_{\sigma_{\tilde{\mathbf{B}}}}= \pi+\langle
\sigma_{\tilde{\mathbf{B}}}|s_z|\sigma_{\tilde{\mathbf{B}}}\rangle=
-\sigma_{\tilde{\mathbf{B}}}\Omega(\tilde{\theta})/2$,  up to $2\pi
n$, where $\Omega(\tilde{\theta})=2\pi(1-\cos\tilde{\theta})$ is the
solid angle subtended by the direction of $\mathbf{s}$.

The dynamical phase vanishes simultaneously in each basis when
\begin{equation}
B_0=[-\frac{\omega}{\kappa} \pm\sqrt{(\frac{\omega}{\kappa}
)^2-4B_1^2}]/2. \label{cond} \end{equation} Then the unitary
transformation becomes purely based on Aharonov-Anandan phases,
$U_{1,aa}=diag(e^{i\pi(B_0+\frac{\omega}{\kappa}
)/\sqrt{(B_0+\frac{\omega}{\kappa} )^2+B_1^2}},
e^{-i\pi(B_0+\frac{\omega}{\kappa}
)/\sqrt{(B_0+\frac{\omega}{\kappa} )^2+B_1^2}}), $ where a global
phase $\pi$ has been omitted.

Now we represent the qubit basis state as
$|\sigma_{\tilde{\mathbf{B}}}\rangle$.   $U_{1,aa}$ becomes $\pi/8$
gate when $(B_0+\frac{\omega}{\kappa} )/|B_1| = -1/\sqrt{63}$, in
addition to condition (\ref{cond}).

The Hadamard gate based on Aharonov-Anandan phases is constructed as
follows. The qubit basis should always be  along the direction
$(B_1,0,B_0+\frac{\omega}{\kappa} )$. The direction of
$\tilde{\mathbf{B}}$  is rotated from
$(B_1,0,B_0+\frac{\omega}{\kappa} )$  by an angle $\chi$ with
respect to $y$ axis, as physically realized by rotating $\mathbf{B}$
from  $(B_1,0,B_0)$ in the same manner. Under condition
(\ref{cond}), one obtains purely geometric gate $U_{1,aa}$, written
in the basis along $\tilde{\mathbf{B}}(0)$. In the qubit basis,
$U_{1,aa}$ if of the form of (\ref{had}), now with
$\gamma=\pi(B_0+\frac{\omega}{\kappa}
)/\sqrt{(B_0+\frac{\omega}{\kappa} )^2+B_1^2}$. Up to a global
factor $i$, the Hadamard gate is realized if $\chi=\pi/4$,
$\gamma=\pi/2$. This value of $\gamma$ is obtained if
$(B_0+\frac{\omega}{\kappa})/|B_1|=1/\sqrt{3}$ is also satisfied.

In the constructions of single-qubit geometric gates, the dynamical
phases in each cycle is proportional to the integral of $B$ or
$\tilde{B}$. Hence the random fluctuation in them integrates to
zero. Besides, the reversal of the sign of the coupling energy in
the second cycle also leads to some cancelation of fluctuations
between the two cycles. Cancelation of the dynamical phase using
spin echoes has been nicely demonstrated
experimentally~\cite{suter,jones,leek}.

Now we turn to two Heisenberg-coupled spins, $\alpha$ and $\beta$,
in the rotating magnetic field, with the Hamiltonian
\begin{equation} {\cal H} = h_{\alpha}+h_{\beta}+J\mathbf{s}_{\alpha}
\cdot\mathbf{s}_{\beta} , \label{ham} \end{equation} where
$h_{j}=-{\kappa}_{j} \hat{\mathbf{s}}_j\cdot\mathbf{B}(t)$,
$(j=\alpha,\beta)$. First we study the two-qubit Berry phases in the
adiabatic limit. Noting that ${\cal H}$ conserves
$\sigma_{\mathbf{B}(t)}^{\alpha}+\sigma_{\mathbf{B}(t)}^{\beta}$, it
is easy to find the instantaneous eigenstates  to be
$|\xi_1(t)\rangle =|\uparrow_{\mathbf{B}(t)}^{\alpha}
\uparrow_{\mathbf{B}(t)}^{\beta}\rangle$, $|\xi_{2/3}(t)\rangle
=\frac{1}{{\cal N}_{\pm}}\{
[-(\kappa_{\alpha}-\kappa_{\beta})\frac{B}{J}
\pm\sqrt{(\kappa_{\alpha}-\kappa_{\beta})^2
\frac{B^2}{J^2}+1}]|\uparrow_{\mathbf{B}(t)}^{\alpha}
\downarrow_{\mathbf{B}(t)}^{\beta}\rangle+
|\downarrow_{\mathbf{B}(t)}^{\alpha}
\uparrow_{\mathbf{B}(t)}^{\beta}\rangle\}$, where ${\cal N}_{\pm}$
is normalization constant, $|\xi_4(t)\rangle=
|\downarrow_{\mathbf{B}(t)}^{\alpha}
\downarrow_{\mathbf{B}(t)}^{\beta}\rangle$. The corresponding
eigenvalues are
$E_1=-\frac{1}{2}(\kappa_{\alpha}+\kappa_{\beta})B+\frac{1}{4}J$,
$E_{2/3}=-\frac{J}{4}\pm
\frac{1}{2}\sqrt{(\kappa_{\alpha}-\kappa_{\beta})^2B^2+J^2}$,
$E_4=\frac{1}{2}(\kappa_{\alpha}+\kappa_{\beta})B+\frac{J}{4}$. The
eigenvalues are time-independent, while the eigenstates are
time-dependent.

When $\phi\rightarrow \phi+2\pi$, $|\xi_i(0)\rangle \rightarrow
e^{i(\gamma_{i}^d+\gamma_{i}^b)}|\xi_{i}(0)\rangle$, where
$\gamma_{i}^d=-E_{i}\tau$ is the dynamical phase, $\tau$ is the
cycling time. $\gamma^b_{i} = i \int_0^{2\pi} d\phi \langle
\xi_{i}(\phi)|\partial_{\phi} \xi_{i}(\phi)\rangle$ is the Berry
phase. It can be found that $\gamma^b_1= \frac{2\pi
B_0}{\sqrt{B_0^2+B_1^2}},$ $\gamma^b_2 =\gamma^b_3 =0$,
$\gamma^b_4=-\frac{2\pi B_0}{\sqrt{B_0^2+B_1^2}}$. They are exactly
equal to $-(s^{\alpha}_{\mathbf{B}(t)}+ s^{\beta}_{\mathbf{B}(t)})$
multiplied by the solid angle $\Omega(\theta)$, up to $2\pi n$.
Therefore for an arbitrary initial state, the unitary transformation
in a cycle  can be represented as
$U_{2,ad}=diag(e^{-iE_1\tau+i\Omega(\theta)},
e^{-iE_2\tau},e^{-iE_3\tau}, e^{-iE_4\tau-i\Omega(\theta)})$, in the
basis $\{ |\xi_1(0)\rangle, |\xi_2(0)\rangle, |\xi_3(0)\rangle,
|\xi_4(0)\rangle \}$.

This result implies that using the Hamiltonian (\ref{ham}), it is
impossible to realize a nontrivial two-qubit unitary transformation
based on Berry phases, which should not be a product of two
one-qubit operations. Even though it can be implemented, any scheme
realizing geometric two-qubit unitary transformations must
ultimately be based on repeated use of the geometric part of $U_{2,
ad}$, i.e.,  $diag(e^{i\Omega(\theta)}, 1,1, e^{-i\Omega(\theta)}),$
which retains the same form in the qubit basis
$\{|\uparrow_{\mathbf{B}(0)}\uparrow_{\mathbf{B}(0)}\rangle,
|\uparrow_{\mathbf{B}(0)}\downarrow_{\mathbf{B}(0)}\rangle,
|\downarrow_{\mathbf{B}(0)}\uparrow_{\mathbf{B}(0)}\rangle,
|\downarrow_{\mathbf{B}(0)}\downarrow_{\mathbf{B}(0)}\rangle\}$, and
can thus be factorized as $(e^{i\Omega(\theta)/2},
e^{-i\Omega(\theta)/2})\otimes(e^{i\Omega(\theta)/2},
e^{-i\Omega(\theta)/2})$, which is but a product of single-qubit
transformations.

We now discuss two-qubit Aharonov-Anandan phases based on the
Hamiltonian (\ref{ham}), with $\phi=\omega t$ specified, but without
the necessity of  adiabaticity. The state is given by
\begin{equation} |\Psi(t)\rangle =
e^{-i\omega t(s^{\alpha}_z+s^{\beta}_z)} e^{-i\tilde{\cal
H}t}|\Psi(0)\rangle,\label{psit}
\end{equation} where $\tilde{\cal H}=
 \tilde{h}_{\alpha}+\tilde{h}_{\beta}+J\mathbf{s}_{\alpha}
\cdot\mathbf{s}_{\beta}, $  with $h_{j}={\kappa}_{j}
\hat{\mathbf{s}}_j\cdot\tilde{\mathbf{B}}_j$,
$\tilde{\mathbf{B}}_j=\mathbf{B}-(\omega/\kappa_j)s^j_z\mathbf{z_1}$,
where $\mathbf{z_1}$ is the unit vector in $z$ direction,
$j=\alpha,\beta$.

When $\kappa_{\alpha}=\kappa_{\beta}=\kappa$, it is impossible to
construct nontrivial two-qubit gates based on Aharonov-Anandan
phases. In this case, the basis states for the two qubits are along
the same direction
$\tilde{\mathbf{B}}_{\alpha}=\tilde{\mathbf{B}}_{\beta}=\tilde{\mathbf{B}}
=(B_1,0,B_0+\omega/\kappa)$. The  eigenstates $|\eta_i\rangle$ and
the eigenvalues $\tilde{E}_i$ of $\tilde{\cal H}$ are in similar
forms to the instantaneous ones of ${\cal H}(t)$ given above, but
with $\mathbf{B}(t)$ replaced as $\tilde{\mathbf{B}}$ and with
$\phi$ replaced as $0$. Starting with $|\eta_i\rangle$, the state at
$t$ is $|\psi_i(t)\rangle= e^{-i\tilde{E}_i t}e^{-i\omega t
(s^{\alpha}_z+s^{\beta}_z)} |\eta_i\rangle, $ which cycles with
period $2\tau=4\pi/\omega$, with the total phase
$\gamma_i=-2\tilde{E}_i \tau$. The dynamical phase is $\gamma^d_{i}=
-\int \langle \psi_i(t)|{\cal H}|\psi_i(t)\rangle dt=
-2\tilde{E}_i\tau-4\pi\eta_i|(s^{\alpha}_z+s^{\beta}_z)|\eta_i\rangle$.
The Aharonov-Anandan phase is
$\gamma^{A}_i=4\pi\eta_i|(s^{\alpha}_z+s^{\beta}_z)|\eta_i\rangle$.
Therefore, $\gamma_1^{aa}=
\frac{4\pi(B_0+\omega/\kappa)}{\sqrt{(B_0+\omega/\kappa)^2+B_1^2}},$
$\gamma_2^{aa}=\gamma_3^{aa}=0$,   $\gamma^{aa}_4=
-\frac{4\pi(B_0+\omega/\kappa)}{\sqrt{(B_0+\omega/\kappa)^2+B_1^2}}.$
Each Aharonov-Anandan phase is
$-(s^{\alpha}_{\tilde{\mathbf{B}}}+s^{\beta}_{\tilde{\mathbf{B}}})$
multiplied by the solid angle $\Omega(\tilde{\theta}),$ up to $2\pi
n$, subtended by a closed curve traced by the direction of the total
spin.  The cycling period is $2\tau$ instead of $\tau$ because
$|\eta_i\rangle$ is an eigenstate
$e^{i2\pi(s^{\alpha}_z+s^{\beta}_z)}$ instead of
$e^{i\pi(s^{\alpha}_z+s^{\beta}_z)}$.  The unitary transformation in
a cycle is $diag(e^{-i2\tilde{E}_1\tau+i2\Omega(\theta)},
e^{-i2\tilde{E}_2\tau},e^{-i\tilde{E}_3\tau},
e^{-i2\tilde{E}_4\tau-i2\Omega(\theta)})$ , in the basis $\{
|\eta_1\rangle, |\eta_2\rangle, |\eta_3\rangle, |\eta_4\rangle \}$.
In the qubit basis, the geometric part is $(e^{i\Omega(\theta)},
e^{-i\Omega(\theta)})\otimes(e^{i\Omega(\theta)},
e^{-i\Omega(\theta)})$, which is a product of single-qubit
operations.

If $\kappa_{\alpha}\neq \kappa_{\beta}$,  the Hamiltonian conserves
neither $s_{\tilde{\mathbf{B}}_{\alpha}}^{\alpha}+
s_{\tilde{\mathbf{B}}_{\beta}}^{\beta}$ nor $s_z^{\alpha}+
s_z^{\beta}$, consequently the geometric phase of each eigenstate
depends on $J$, the geometric part of the unitary transformation is
not simply a product of geometric unitary transformations of the two
qubits. Therefore, nontrivial two-qubit geometric gates are allowed
in principle. When $\kappa_{\alpha}\neq \kappa_{\beta}$ while $J$ is
much smaller than $|\kappa_{\alpha}|$ and $|\kappa_{\beta}|$, it is
possible to choose $\omega\approx \kappa_{\alpha}B_0$ so that it
becomes very reasonable to make the approximation that $\beta$ spin
is decoupled with the field. Under the same condition, Heisenberg
interaction in liquid NMR can be approximated as Ising coupling
$Js_z^{\alpha}s_z^{\beta}$.  Therefore, previous treatment of the
two-qubit Aharonov-Anandan phases~\cite{jones,zhu,wang} is fully
justified.

For scalable quantum computing, it may be more convenient to use
qubits with $\kappa_{\alpha}=\kappa_{\beta}$, while the systematic
error due to approximations in each gating may accumulate after many
runs of many gates~\cite{shi}.  Therefore, one may adopt a hybrid
scheme combining geometric one-qubit gates and dynamical qubit
gates, e.g. $\sqrt{\rm SWAP}$,  based solely on Heisenberg
interaction. It is well known that ${\rm
CNOT}=(T_{\alpha})^2(T_{\beta}^{-1})^2 \sqrt{\rm
SWAP}(T_{\alpha})^4\sqrt{\rm SWAP}$, where the $\pi/8$ gate $T$ can
be constructed by using Berry or Aharonov-Anandan phases. This
hybridization is quite natural, because the isotropy of Heisenberg
interaction allows $\sqrt{\rm SWAP}$ to be realizable under our
representation of qubit basis states.

To summarize, we have examined the implementation of geometric
quantum computation based on Zeeman coupling with a rotating
magnetic field and isotropic Heisenberg interaction. Using a novel
representation of  the qubit basis states, we have constructed the
standard $\pi/8$ and Hadamard single-qubit gates in terms of Berry
and Aharonov-Anandan phases. In principle, these constructions of
geometric single-qubit gates can be adapted to all implementations
of quantum computing. For two Heisenberg-coupled qubits in the
rotating field, we show that unitary transformations based on Berry
phases, as well as those based on Aharonov-Anandan phases in case
the gyromagnetic ratios of the the two qubits are equal, are all
products of single-qubit transformations. Hence nontrivial two-qubit
geometric gates in these cases are impossible. But two-qubit
transformations based on Aharonov-Anandan phases  are nontrivial in
general when the two gyromagnetic ratios are different, in
consistency with previous works. We suggest that one may adopt a
hybrid approach combing geometric $\pi/8$ and Hadamard gates with a
dynamical $\sqrt{\rm SWAP}$ gate.

I thank Professors Q. Niu, R. B. Tao, Z. D. Wang, B. Wu and S. L.
Zhu for useful discussions. This work is supported by National
Science Foundation of China (Grant No. 10674030),  Pujiang Project
(Grant No. 06PJ14013), Shuguang Project (Grant No. 07S402), and
partly by PKIP of CAS.

\end{document}